# Variable Instruction Fetch Rate to Reduce Control Dependent Penalties

Aswin Ramachandran, PhD, Oklahoma State University

Louis G. Johnson, Emeritus Professor, Oklahoma State University

http://www.ece.okstate.edu/content/johnson-louis-g-phd

***Abstract*—In order to overcome the branch execution penalties of hard-to-predict instruction branches, two new instruction fetch micro-architectural methods are proposed in this paper. In addition, to compare performance of the two proposed methods, different instruction fetch policy schemes of existing multi-branch path architectures are evaluated. An improvement in Instructions Per Cycle (IPC) of 29.4% on average over single-thread execution with gshare branch predictor on SPEC 2000/2006 benchmark is shown. In this paper, wide pipeline machines are simulated for evaluation purposes. The methods discussed in this paper can be extended to High Performance Scientific Computing needs, if the demands of IPC improvement are far more critical than $cost.**

***Keywords*—IPC, Branch Prediction, Multi-branch Path Executions, cycle-accurate simulation, Instruction-Level Parallelism, SMT processors, Superscalar, confidence estimator, SPEC benchmarks, High Performance Computing.**

## I Introduction

Control dependencies in a program can be related indirectly to data dependencies. Nevertheless, the control flows of the program seem to be predicted to a fair degree of accuracy (Nair, 1995 [1]) for machines with small instruction fetch. But, it introduces a limitation for wider instruction fetch machines and is harder to predict the control flow. This is because of lack of sophisticated hardware with small latency to recognize the pattern of the program behavior or in general, due to the innate behavior of the program.

### A. Higher IPC with Superscalars

Figure 1 compares the fraction of branch misprediction that have a probability of error more than 0.3 and between 0.3 and 0.7. As seen from the plot, about 45% of branches are mispredicted. Even if the predictions of the branches that have a probability of error greater than 0.7 are overridden (since there are wrongly correlated (Klauser, 2001, [2])), there are still about 38% of the branches whose behavior patterns are not correlated with the branch predictor.

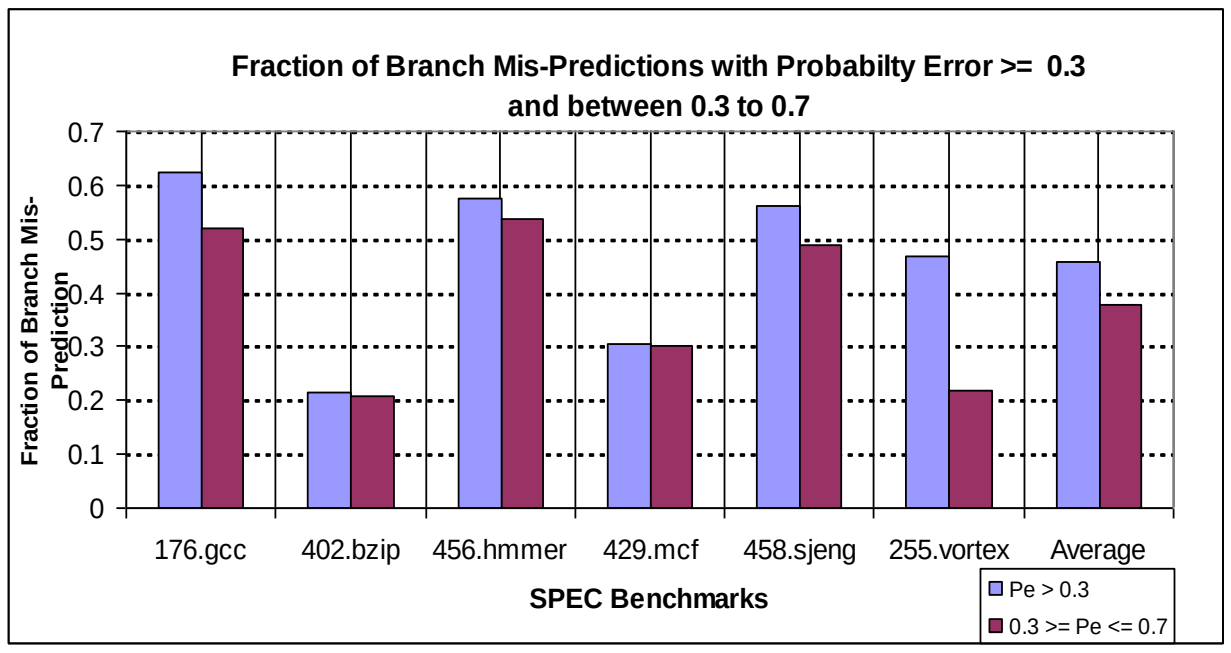

Fig. 1. Fraction of Branch Misprediction in SPEC CPU INT 2000/2006 benchmarks (gshare: Size: 2048 entries; History Bits: 16; BTB: 512 sets with 4-way associative)

The goal of the superscalar architecture design is exploit available Instruction Level Parallelism (ILP) in the program code and hence, to achieve maximum IPC. But, to maximize the utilization through ILP, the control flow of the program has to be predicted with good accuracy. As the instruction fetch width increases the number of branches in the fetch group also increases. The branch predictor has to now choose among multiple branch paths and predict the correct path. The problem is exacerbated as the machine is super-pipelined. Due to the increase in branch execution latencies, there are more unresolved pending branches. This paper evaluates multi-path schemes where both the taken and not taken branch paths are followed and executed speculatively.

The following are the contributions of this paper,

- Propose and evaluate two new disjoint-eager execution schemes
  - Selective disjoint-eager execution.
  - Dynamic disjoint-eager execution.
- Propose simple hardware design logic to create, manage and destroy speculative eager threads.
- Finally, evaluate the performance of eager execution schemes.

## II. Related Work

Ahuja et al, 1998 [3] show average speedups of 14.4% for multipath architecture with confidence predictor on SPECint95 benchmarks compared to a single path machine. The paper demonstrates that the instruction fetch bandwidth is very important and extra resources to fetch correct execution path can improve performance. However, the study does not indicate how the fetch resources must be allocated and how the confidence values can be used to control the fetch allocation.

JRS confidence estimator by Jacobsen, Rotenberg and Smith, 1996 [4] introduce the concept of confidence estimators. They test the performance of confidence estimator with ones counter (shift registers), saturating and resetting counter. The paper shows that resetting counter tracks ideal curve of misprediction due to dynamic branches closely than other counter methods. Selective Branch Inversion (SBI) is proposed by Klausaur et al., 2001, [2]. An up-down counter is used in the confidence estimator with 0 marked as low confidence and 1 to 3 as high confidence. A relative improvement of 9% reduction in branch misprediction is noted when compared with the McFarling predictor. However, performance improvement in terms of IPC is not indicated in the paper. Manne et al 1999 [6] also introduces various useful confidence evaluation metrics such as PVN and Specificity.

Uht et al., 1995 [7] propose a variation in eager execution schemes called the Dis-Joint Eager Execution (DEE). It uses the cumulative path probabilities to determine the highest likelihood path to follow. A mean speedup of 4% over single path execution if more than 256 possible paths are followed is recorded. However, the implementation of DEE is simplified by only considering the static branch prediction probabilities and does not consider the dynamic probabilities for each individual branch. In addition, the paper also does not propose any realistic hardware design to implement DEE.

Dual Path Instruction Processing is proposed by Aragon et al, 2001 [5] using Branch Prediction Reversal Unit (BPRU). This architecture targets to reduce the pipeline-fill penalty after a misprediction. An 8% improvement is noted over single path with gshare predictors. However, fetching from alternative streams reduces the fetch bandwidth and more than 2 branch paths have to be followed as shown in DEE.

Selective Dual Path with various fetch polices using confidence values is studied by Heil and Smith, 1997 [8]. The fetch policies did not provide much improvement and the paper concludes to investigate on machines that can fork multiple branch paths. Wallace et al., 1998 [9] propose a method to use the 2-way SMT for multipath execution. They use a fetch policy called the ICOUNT, where the fetch logic gives priority to those threads that have fewest instructions between fetch and issue. A 14 % increase in this modified SMT over the baseline architecture is seen.

## III. Design Approach

The multi-path design using some form confidence estimators has been proposed earlier. Klauser et al., 2001 [2] discuss about Selective Eager Execution using confidence estimator and achieve an average improvement of 14% in IPC for SPECint95 benchmarks. However, schemes such as the DEE (Uht and Sindagi, 1995 [7]) have never been evaluated with realistic architecture designs and with dynamic confidence estimators. The performance improvement varies from 4% to 14% in most of the architecture designs that tried to improve the single-threaded program execution. In addition, the performance of the multi-path design relies to an extent on the performance of the dynamic confidence estimators. In the following sections, the fetch policies and design aspects of the SMT architecture are explained in detail.

### A. Selective Disjoint-eager Execution

In DEE [7], the instructions are fetched from the path that has the highest path confidence. However, dynamic confidence estimators are shown to have problems due to aliasing and difficultness to measure the predictor and the branch behavior. In this paper, to alleviate the inaccuracies of the confidence estimator, a set of thread paths are followed. We use the term thread path to highlight the point that the paths have separate registers and execute in parallel.

For example, if the fetch width is 32 instructions per cycle and the desired IPC is at least 8, then 4 thread paths each of 8 instructions that have high confidence are fetched in a single cycle. Basically, this scheme follows the set of paths that have high likelihood to be correct and controls the over bound growth of thread paths in the eager execution scheme.

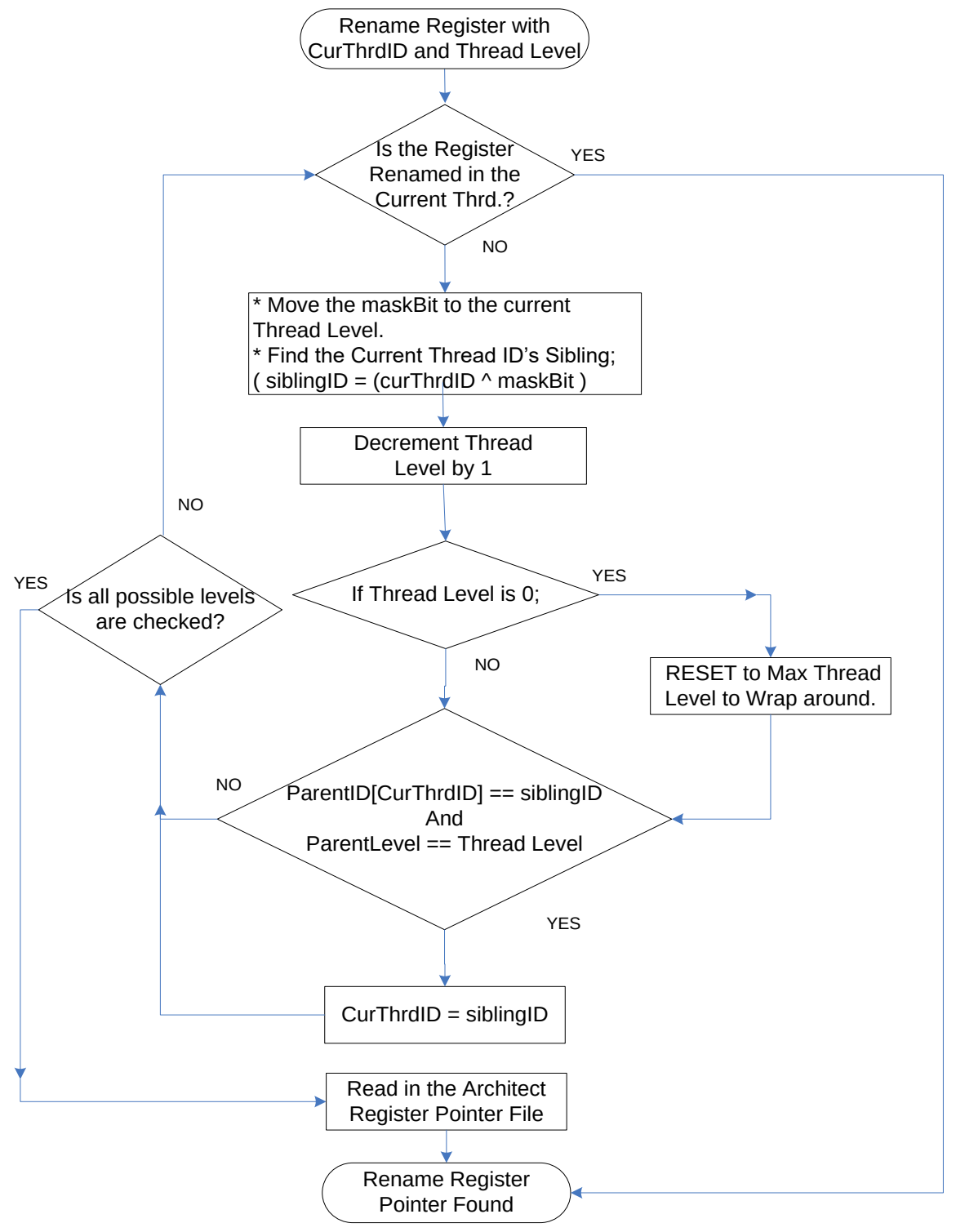

Flow Chart 1. Thread Rename Pointer Logic

## B. Dynamic Dis-Joint Eager Execution

This scheme is similar to DEE with selective paths except that instead of a fetching a fixed number of instructions per thread path, the number of instructions fetched are proportional to the confidence value of that path.

For example, in a 32-wide fetch machine if the confidence values of high confidence threads, Thread1 and Thread2 are 0.8 and 0.2, respectively. Then, about 26 instructions will be fetched for Thread1 and 6 instructions for Thread2 in a particular cycle. As the confidence values changes the fetch bandwidth for the threads also changes proportionally. To solve the problem of misprediction penalties in single-thread instruction stream, a scheme were multiple paths are followed and executed using Simultaneous Multi-Threaded (SMT) architecture designs is adapted.

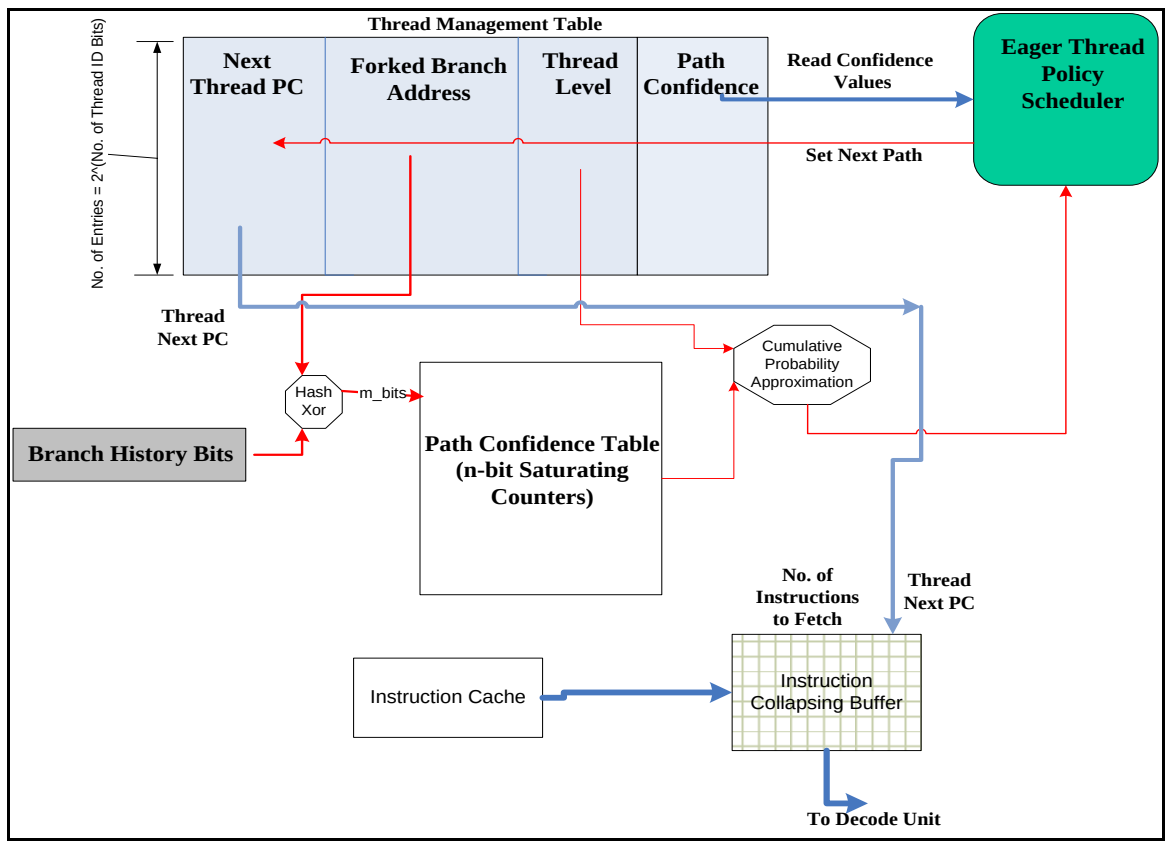

Fig. 2. Logical Block Diagram of Fetch Policy using Confidence Estimator

In addition, several policies can be applied to choose the set of maximum likelihood thread paths and are explained in Table 1. It also increases the utilization of fetch resources rather than just following one high confidence path or following all possible paths.

## C. Multipath Fetch Logic Design

The instruction fetch scheduler may use different fetch policies that are listed in Table 1. In the case of the multiple paths, a multi-ported BTB and instruction cache are necessary to determine multiple target addresses. The challenge in fetching from multiple paths is to make sure the instructions from these streams can be distinguished at any point inside the processor. This is could be done in 2 ways. Structurally the entire processor can be divided for each of these streams or each instruction can be tagged with a path or thread identification tag – Thread ID – to distinguish between various paths.

Structurally dividing the entire processor may enforce strict limitation of number of threads and also that these resources can be shared. Hence, to improve resource utilization the hardware functional units and registers must be shared among these paths. Therefore, a unique scheme where the branch history bit is used for Thread IDs is proposed by Chen, 1998 [11]. Through this scheme the taken path is set as 1 and the not taken path is set as 0. The logical block diagram of fetch logic design is shown in Figure 2.

## D. Register Renaming for Multiple Paths

Although, the register renaming mechanism for multi-path architecture is same as for single-threaded out-of-order executions, one major difference in this architecture is that the renaming can happen at any level of the forking path. Hence, the challenge is to find the correct ancestor path and also to reference the correct rename pointer. Let's look at the procedure to find the correct ancestor thread ids through an example.

TABLE 1 COMPARISON OF FETCH POLICY SCHEMES

| Policy | **Single-Thread Execution** | | **Eager Execution** | **Dis-Joint Eager Execution (DEE)** *[* proposed in this paper]* | | |
|---|---|---|---|---|---|---|
| | *Perfect branch prediction* | *gshare branch prediction* | *Divided Fetch Width* | *DEE* | ***Selective DEE** | ***Dynamic DEE (Variable-Fetch Rate)*** |
| Fetch Group | Depends on BTB and perfect predictor | Depends on BTB and Branch Predictor | Split equally among all active paths | Only one path with high confi-dence values is chosen | A *set of paths* with high confidence values with fixed number of instructions is chosen. | Allocated *proportionally* among all paths based on Confidence Values |
| Reason to study this scheme | Perfect case | To prove branch prediction for high fetch band-width is poor. | To illustrate the machine performance without any kind of branch prediction | To limit the number of threads with confidence values | To minimize of dependence on confidence values as they can be misleading | To enable instruction fetch for each path, only restricted by its confidence value and the total fetch width. |
| Max. Possible Number of Threads | 1 | 1 | $2^n$, where 'n' is no. of branch levels | $2^n$, where 'n' is no. of branch levels | Depends on the Target IPC limit | $2^n$, where 'n' is no. of branch levels |
| Conditional Branch Prediction | Perfect | 2-Bit State Predictor | Used after maximum thread level | Used after maximum thread level | Used after maximum thread level | Used after maximum thread level |
| Confidence Estimator | No | No | No | Yes | Yes | Yes |
| Additional Hardware | NA | Counters and BTB | Multi-Path machine | Confi-dence multi-pliers | Confidence multipliers & priority encoder. | Multipliers for proportional allocation at Fetch |

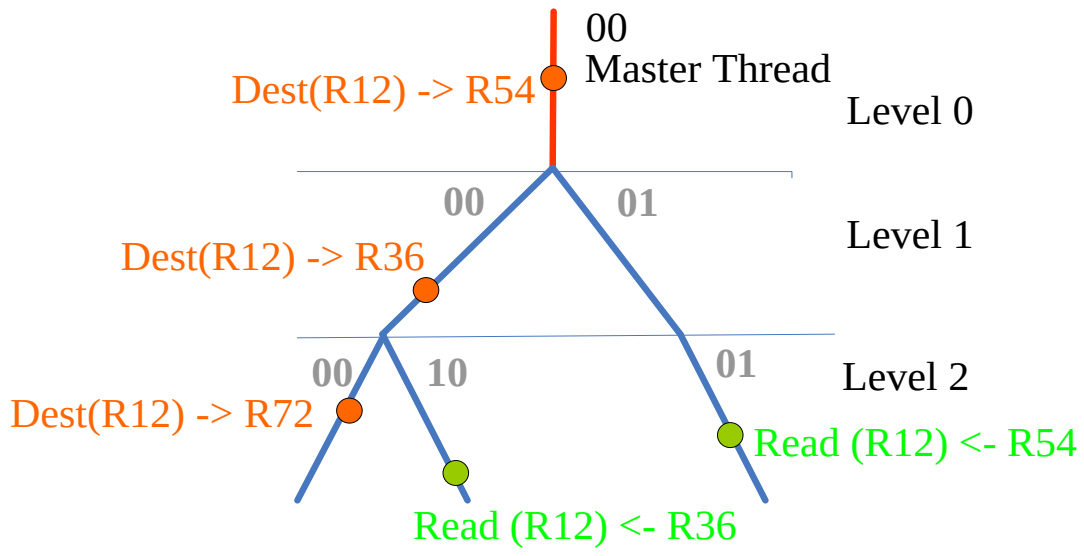

Fig, 3. Example of Register Renaming in Multi-Path Design

In the example shown in Figure 3, register 12 gets renamed once at the master thread as well as twice in Thread ID 00 but at different branch levels. In thread paths 10 and 01, register 12 is being read and the correct register pointers are indicated by arrow symbols in the Figure 3. The explanation of how register 12 references correctly to its renamed pointers is given in the Flow Chart 1.

Rename register logic is one major module that different from that of single-path architecture design. The rest of the units in the pipeline in the multi-path architecture design are similar to single-path. However, to reduce the number of thread paths that are followed, the thread paths are invalidated at dispatch and complete stages as soon as the branch get executed and its actual path is determined. The reason to keep the number of thread path low in a multi-path scheme is because the more the number of thread paths that are followed the less is the fetch width per thread.

## IV Simulation Environment

AbaKus simulation framework is used to explore the architectural features with both the branch prediction and multi-path execution schemes. This framework with module and port-structures gives a fair degree of accuracy in the simulations with reasonable speed. The details of AbaKus framework and superscalar models are discussed [10].

To focus the study on conditional branch effects on the processor, the component designs of simulated architecture are widened to minimize any structural design hazards. Perfect memory is assumed as conditional branches only have indirect effect on memory. The summary of architecture details are described in Table 2. The simulation is executed using Intel Xeon CPU 3.2 GHz (128-node cluster) with 4GB RAM. In the next section, the architecture descriptions of the single-threaded and multi-threaded designs are discussed.

TABLE 2 Simulation Details of the Multi-Path SMT Architecture

| **Design Parameters** | **Multi-Path SMT** |
|---|---|
| Maximum No. of Threads | $2^{25}$ possible threads. Exclusively depends on Fetch Policy |
| Instruction Fetch Width per Thread | 8 or 32 insts/cycle but depends on fetch policy |
| Instruction Window Size | 4096 entries |
| Physical Registers | 32 |
| Issue Width | 64 |
| Commit Width | 128 |
| BTB & Branch Predictor (if used) | BTB: 8192 16-way Gshare: 16384 entries; 16 History Bits |
| Confidence Estimator (if used) | 8132 entries |
| Confidence Counters (if used) | 4-bit Saturating Counters |
| Integer ALU units (Latency =1) | 40 |
| Branch Units (Latency = 1) | 40 |
| Load/Store (Latency = 2) | 40 |
| Mul/Div (Latency = 5) | 20 |
| Float/Special Units (Latency = 3) | 40 |
| Write Back Bus Width | 128 |
| Complete Width | 128 |

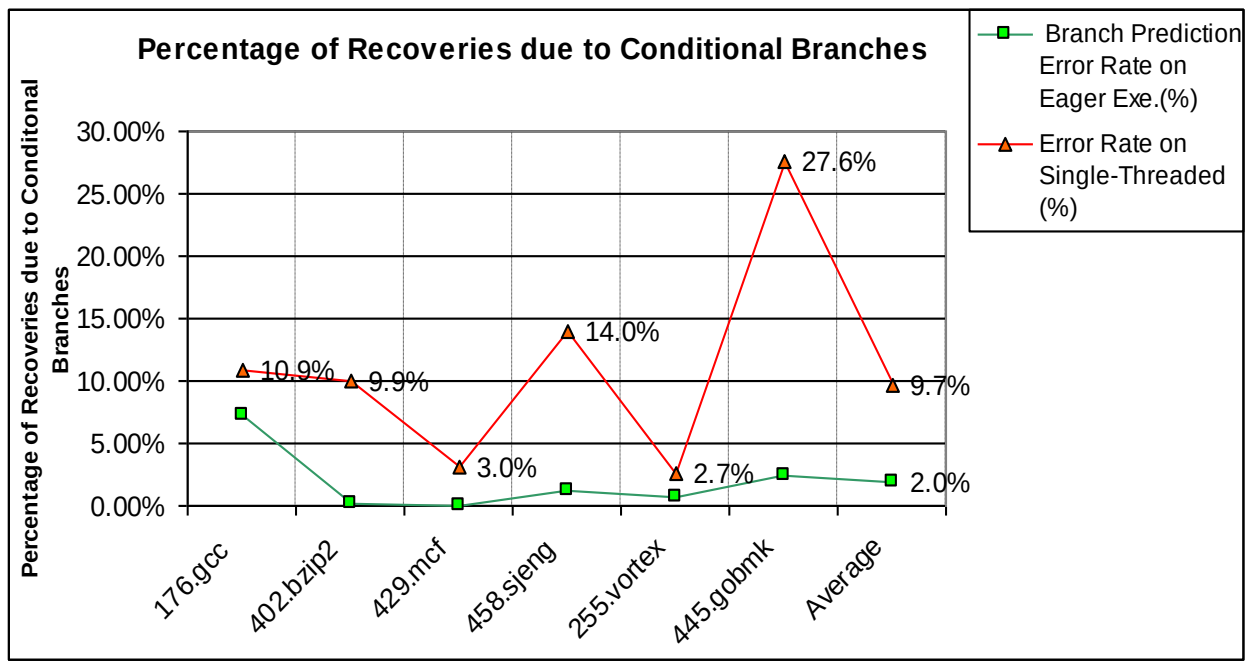

Fig, 4. Percentage of Recoveries due to conditional branch misprediction

## V Design Implications

The benchmarks are run up to 500 million instructions and then the architecture designs are tested for the next 100 million instructions. This set of 100 million instructions, however, does not represent the entire benchmark that typically has more than 1 trillion instructions.

To understand the performance limitations of the conditional branches, a processor with perfect conditional branches is evaluated. This is done by gathering the target address traces of the conditional branches in a single-threaded processor and then, allowing the simulation to read from this trace when a conditional branch is encountered. In this way all the architecture parameters are the same between the perfect and the single-thread processor except the conditional branch prediction.

### A Confidence Estimator

Another approach to reduce the number of paths is to follow the path that has the most likelihood to be executed. This form of execution is called Dis-Joint Eager Execution (DEE) and is discussed in detail Table I. In this section, the design and performance of the confidence estimator is discussed. The performance of the 4-bit saturating confidence counter and other performance metrics are discussed by Manne et al., 1998 [38].

The following is the Pseudo-Code of the confidence update mechanism when the branch executes:

```
Prediction Correct:
  if confidence value < 8 (Low Confidence) Then
     Set confidence value = 8
  if confidence value >= 8 (High Confidence) Then Increment
confidence value

Prediction Incorrect:
  if confidence value < 8 (Low Confidence) Then
Increment confidence value
  if confidence value >= 8 (High Confidence) Then
     Set confidence value = 7
```

The major difference with fetching instructions based on confidence estimates is that instead of using a branch predictor, a table of saturating counters is used by the fetch scheduler to determine the path of the next instruction fetch. The branch target buffer is now augmented by Thread Management Table.

The Thread Management Table has the following fields, the next Thread PC, the forked branch address, thread level and path confidence. These fields are explained below,

*Next Thread PC:* Stores the next program counter of each active path.
*Forked Branch Address:* This is the branch address where the path is forked.
*Thread Level:* Indicates the level of the thread path.
*Path Confidence:* Stores the confidence value of the path.

### B. Reducing Conditional Branch Mispredictions

The eager-based fetch policy schemes are detailed in Table 1. Figure 4 shows the percentage of recoveries due to conditional branch mispredictions for multi-path eager execution policy and single-threaded branch predictions. Figure 4 show that eager execution has reduced the number of recoveries. Mispredictions in eager based executions are due to compulsory BTB misses and if the number of unresolved branches reaches the maximum number of branch levels possible in the processor.

Branch prediction is used in the eager-based execution only if the maximum possible unresolved branch level is reached in the processor. If branch prediction is used then it leads to a possibility of misprediction. Hence, it is important for eager-based executions to use branch prediction rarely by increasing the number of maximum possible branch levels in the machine. This results in increase in more possible threads to handle in the processor. For example, if 3 unresolved branches exist in the processor then it leads to a maximum possibility of 23 or 8 threads. The results of the simulations with IPC as the measure of performance for 32-wide fetch are shown in Figure 6.

One subtle but important observation is that dynamic confidence estimator performs well than just having static confidence estimator. This is illustrated in Figure 6, as the DEE with static confidence performs poorly than the single-threaded execution.

For 32-wide fetch machine, the maximum possible improvement between the processor with perfect conditional branch prediction and the single-threaded processor with gshare branch prediction is about 70% on average. 0.99.go has the best improvement on IPC with about 77.26% for the 32-wide fetch with eager execution. On average, the eager execution shows 29.44 % improvement over single-threaded execution with branch prediction.

Eager polices that depend on confidence values assumes that branch prediction error can be mitigated by using confidence estimates. But, given the inaccuracies of confidence estimates, the dynamic DEE (IPC=1.58) and selective DEE (IPC=1.51) have less IPC than the eager execution policy (IPC=1.72).

Using the confidence estimator described by Manne et al, 1998 [6] only supplements branch prediction. In addition, the dynamic nature of code execution proves to be far more complex than the confidence estimator can handle. This is illustrated in Figure 5 that shows the values of PVN, PVP, Specificity and Sensitivity of the confidence estimator. It is important that PVN – probability that low confidence is mispredicted correctly and Specificity – fraction of mispredictions that are low confidence are close to 1.

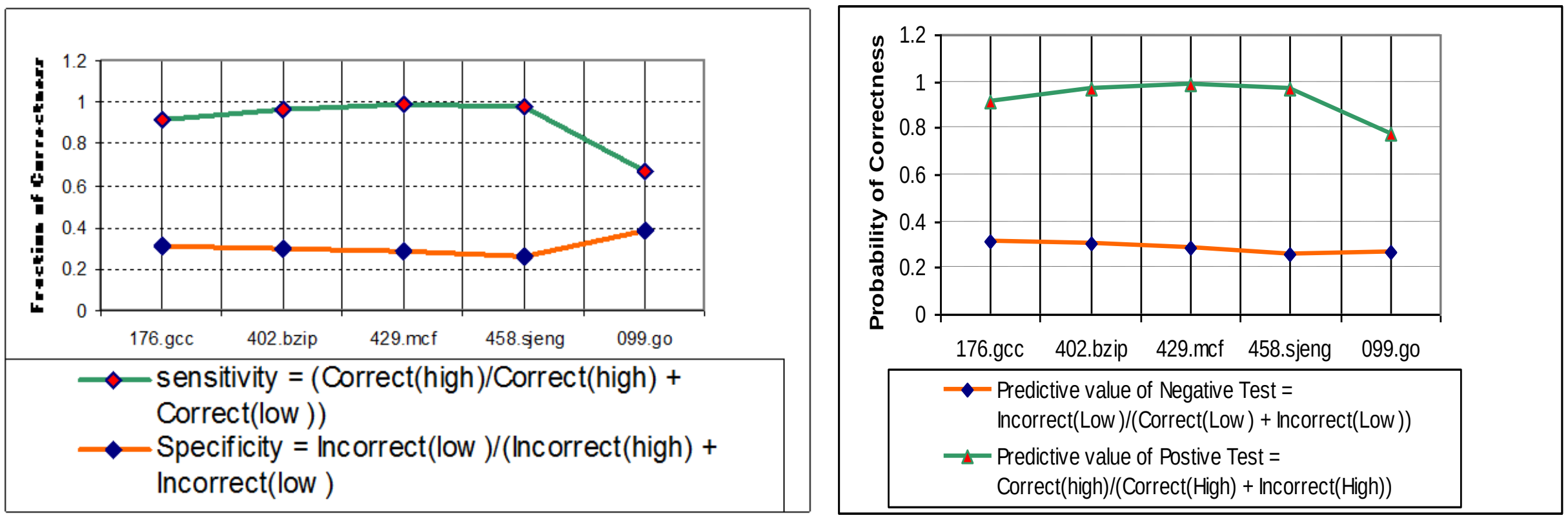

Fig. 5 Accuracy of the Confidence Estimator with 4-bit saturating counters. The low values of PVN and Specificity highly affects the performance of the confidence-based eager executions.

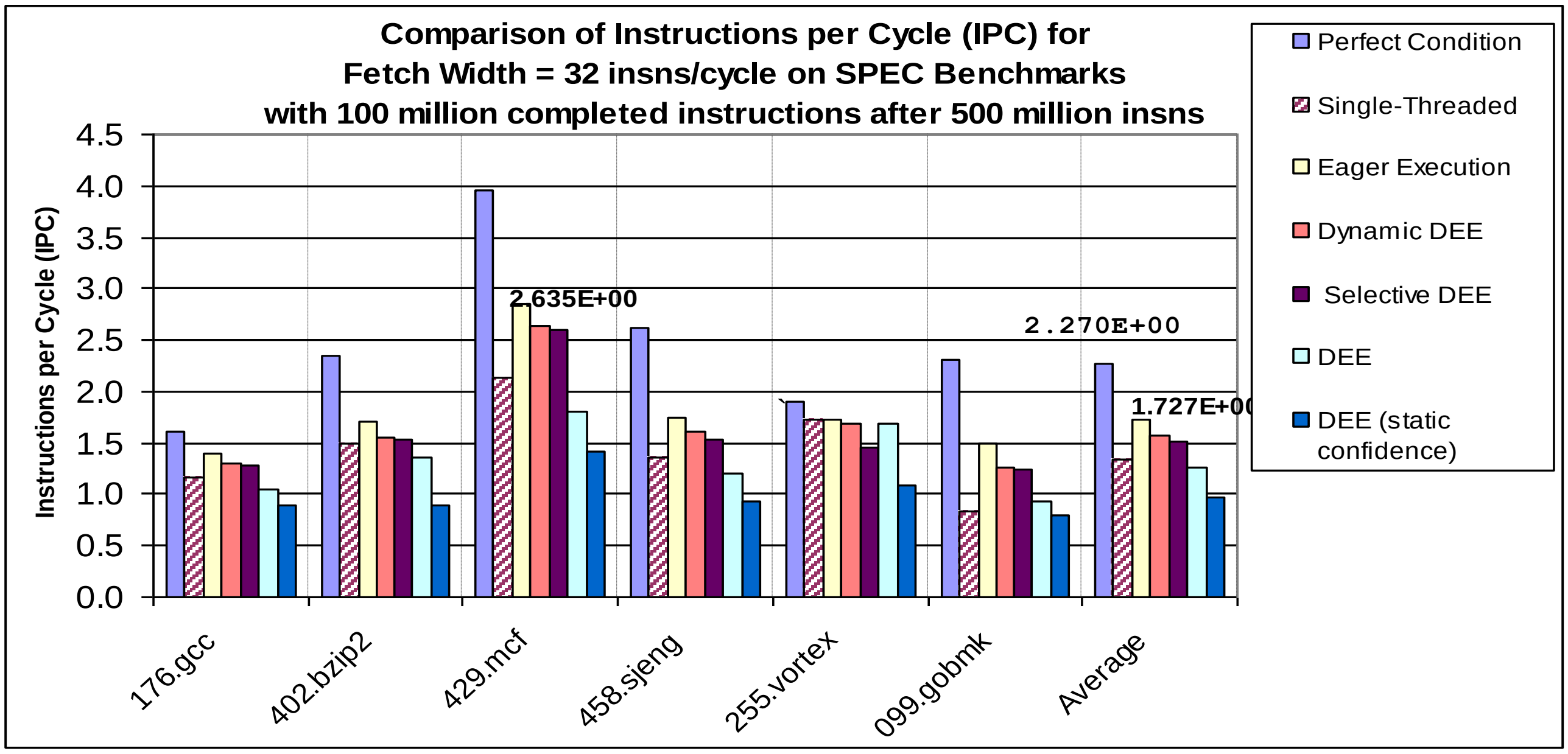

Fig. 6 Comparison of IPC for different eager-based polices with single-threaded processor for 32-wide fetch. **On average, single thread IPC is 1.4 and Eager Execution 1.72 which is about 30% maximum IPC improvement.**

## VI Conclusion and Future Work

Wide-pipeline hypothetical machines are simulated only to evaluate the true limitations of the single-threaded code. There are 3 important factors that need to be considered to attain the IPC of the perfect conditional branch prediction – confidence estimates, branch prediction and fetch width. In addition to confidence estimators, branch prediction and fetch width have a direct effect on IPC. The use of branch prediction is dependent on the maximum number of branch levels available in eager execution schemes. If the eager schemes have more number of branch levels, then the numbers of active threads increases resulting in division of fetch resources.

The way in which the fetch resources are divided depends on the imposed fetch policy of processor. However, as a result of dividing the fetch resources the number of instructions supplied to each thread is reduced impacting the IPC.

The eager and disjoint-eager based executions of 25 and 16 levels have more or less a similar IPC where as the disjoint-eager with 8-levels have less number of threads but falters as it relies more on the branch predictor. The effect on conditional branch misprediction on IPC of the processor is clearly seen in Figure 4. There is about 70% performance loss due to such mispredictions. As the number of available branch levels decrease the processor relies more on the branch predictor and tend to make more branch mispredictions. This directly results in decrease in IPC.

The size of each benchmark (more than 1 trillion instructions) and code phase variations makes it challenging to understand the true performance of the architecture design. However, by statistical and other clustering techniques subsets of code that represents the entire benchmark can be determined. This can help in finding sensitive regions of code snippets to evaluate future eager-execution based architecture designs. The 30% average improvement (Fig. 6) in IPC for eager-based execution over single-threaded execution with branch prediction is significant considering the benchmarks that are chosen for this research in the performance evaluations.

## Acknowledgments

The authors would really like to thank the computer architecture faculty in the School of Electrical Engineering, …. University and Graduate College for the sponsorship of this research work. The authors are also thankful for the … High Performance Computing cloud resources.